\documentclass[11pt]{article}
\usepackage{graphicx}

\usepackage{graphicx}
\usepackage{dcolumn}
\usepackage{amsmath}
\usepackage{units}

\input{pubboard/babarsym}

\newcommand{\Btag}{\ensuremath{\B_\mathrm{tag}}}
\newcommand{\Brec}{\ensuremath{\B_\mathrm{rec}}}
\newcommand{\Bztojpsiks}{\ensuremath{\Bz\to\jpsi\KS}}

\newcommand{\sintwobeta} {\ensuremath{\sin 2\beta}}

\newcommand{\Bztokspiz} {\ensuremath{\Bz \to \KS\piz}}
\newcommand{\Bztokzpiz} {\ensuremath{\Bz \to \Kz\piz}}
\newcommand{\ckspiz} {\ensuremath{C_{\KS\piz}}} \newcommand{\skspiz}
{\ensuremath{S_{\KS\piz}}} \def\cf {\ensuremath{C_f}} \def\sf
{\ensuremath{S_f}} 
 
\newcommand{\Bflav}
{\ensuremath{B_{\rm flav}}} \newcommand{\zrec}{\ensuremath{z_{\CP}}}
\newcommand{\ztag}{\ensuremath{z_\mathrm{tag}}}
\newcommand{\mmiss}{\ensuremath{m_\text{miss}}}
\newcommand{\mb}{\ensuremath{m_{B}}}
\newcommand{\BABARPubYear}    {06}

\newcommand{\BABARConfNumber} {030}
\newcommand{\SLACPubNumber} {11982}

\newcommand{\thetacms}{\ensuremath{\theta_{B}^*}}
\newcommand{\costhetacms}{\ensuremath{\cos\thetacms}}

\def\figurebox#1#2#3{%
    \def\arg{#3}%
    \ifx\arg\empty
    {\hfill\vbox{\hsize#2\hrule\hbox to #2{\vrule\hfill\vbox to #1{\hsize#2\vfill}\vrule}\hrule}\hfill}%
    \else
    {\hfill\epsfbox{#3}\hfill}%
    \fi}

\long\def\inst#1{\par\nobreak\kern 4pt\nobreak
    {\it #1}\par\vskip 10pt plus 3pt minus 3pt}

\begin{document}

{\pagestyle{empty}


\begin{flushright}
\babar-CONF-\BABARPubYear/\BABARConfNumber \\
SLAC-PUB-\SLACPubNumber \\
July 2006 \\
\end{flushright}

\par\vskip 5cm

\begin{center}
\Large \bf   \boldmath Measurement of the \CP-violating Asymmetries
	in \Bztokspiz\ and of the Branching Fraction of \Bztokzpiz\ 
\end{center}
\bigskip

\begin{center}
\large The \babar\ Collaboration\\
\mbox{ }\\
\today
\end{center}
\bigskip \bigskip

\begin{center}
\large \bf Abstract
\end{center}
  We present a measurement of the time-dependent $C\!P$-violating
  asymmetries in $B^0 \to K^0_{\scriptscriptstyle S}\pi^0$\ decays
  based on $348$ million $\Upsilon(4S)\to B\kern 0.18em\overline{\kern
  -0.18em B}$ events collected by the \mbox{\slshape
  B\kern-0.1em{\smaller A}\kern-0.1em B\kern-0.1em{\smaller
  A\kern-0.2em R}}\ experiment at the PEP-II asymmetric-energy $B$
  Factory at SLAC. We measure the direct $C\!P$-violating asymmetry
  $C_{K^0_{\scriptscriptstyle S}\pi^0} = 0.20 \pm 0.16 \pm 0.03$ and
  the CP-violating asymmetry in the interference between mixing and
  decay $S_{K^0_{\scriptscriptstyle S}\pi^0} = 0.33 \pm 0.26 \pm 0.04$
  where the first error is statistical and the second systematic. On
  the same sample, we measure the decay branching fraction, obtaining
  ${\cal B}(B^0 \to K^0_{\scriptscriptstyle S}\pi^0)= (10.5 \pm 0.7
  \pm 0.5)\times 10^{-6}$. All results presented here are preliminary.

\vfill
\begin{center}

Submitted to the 33$^{\rm nd}$ International Conference on High-Energy Physics, ICHEP 06,\\
26 July---2 August 2006, Moscow, Russia.

\end{center}

\vspace{1.0cm}
\begin{center}
{\em Stanford Linear Accelerator Center, Stanford University, 
Stanford, CA 94309} \\ \vspace{0.1cm}\hrule\vspace{0.1cm}
Work supported in part by Department of Energy contract DE-AC03-76SF00515.
\end{center}

\newpage
}

%
\begin{center}
\small

The \babar\ Collaboration,
\bigskip

%
{B.~Aubert,}
{R.~Barate,}
{M.~Bona,}
{D.~Boutigny,}
{F.~Couderc,}
{Y.~Karyotakis,}
{J.~P.~Lees,}
{V.~Poireau,}
{V.~Tisserand,}
{A.~Zghiche}
\inst{Laboratoire de Physique des Particules, IN2P3/CNRS et Universit\'e de Savoie,
 F-74941 Annecy-Le-Vieux, France }
% ...

\end{center}\newpage

The recent measurements of the weak phase $\beta$ 
in $b \to c \bar c s$ decays 
from \babar~\cite{BaBarSin2betaObs} and Belle~\cite{BelleSin2betaObs},
have reached the precision of the prediction from
fits of the unitarity triangle~\cite{fits}, obtained combining
the information from \CP-conserving quantities  to the measurements
of other \CP-violating (CPV) processes. The agreement
between the two determinations has shown
that the Cabibbo-Kobayashi-Maskawa (CKM) 
quark mixing matrix~\cite{CKM} correctly describes the source of
effects in the Standard Model (SM).

With \babar\ and Belle collecting more data, one of the 
major goals of the two experiments is to search for indirect
evidence of new physics (NP). One possible strategy consists in comparing the 
established value of $\beta$
to independent determinations of the same quantity, obtained
from penguin-dominated (in SM) $b \to s\qbar\q$
$(\q=\{d,s\})$ 
decays~\cite{Grossman:1996ke,Ciuchini:1997zp}.~\footnote{Unless 
explicitly stated, conjugate decay modes are
assumed throughout this paper.}

In the SM, the parameters \cf{} (describing  the direct CPV asymmetry) and
\sf{} (describing the CPV asymmetry in the interference between mixing
and decay) are expected to be consistent with the values from $b \to c \bar c s$ decays
(namely \cf{}$\sim 0$ and \sf{}$\sim \sin 2\beta$). Small deviations from
this expectation can be induced by additional CKM suppressed contributions
to the amplitude. On the other hand, additional radiative loop contributions 
from NP processes may produce large deviations.

In this letter we present updated measurements of the time-dependent
CPV asymmetries and branching fraction of the decay \Bztokzpiz.  
The CKM and color suppression of the tree-level $b\to s\bar{u} u$ transition 
leads to the expectation that this decay is dominated by a top quark 
mediated $\b\to\s\dbar\d$ penguin diagram, which carries a weak phase
$\arg(V_{\t\b}V_{\t\s}^*)$. If non-leading contributions
are ignored, the time-dependent CPV asymmetry is governed by \sintwobeta. 

The results presented here are based on $348$ million $\Y4S\to\BB$
decays collected with the \babar\ detector at the PEP-II
$\epem$ collider, located at the Stanford Linear Accelerator Center.
The \babar\ detector, which is described in~\cite{ref:babar},
provides charged particle tracking through a combination of a
five-layer double-sided silicon micro-strip detector (SVT) and a
40-layer central drift chamber (DCH), both operating in a
\unit[1.5]{T} magnetic field to provide momentum
measurements. Charged kaon and pion identification is achieved through
measurements of particle energy loss ($dE/dx$) in the tracking system
and Cherenkov cone angle ($\theta_c$) in a detector of internally
reflected Cherenkov light (DIRC).  A segmented CsI(Tl) electromagnetic
calorimeter (EMC) provides photon detection and electron
identification.  Finally, the instrumented flux return (IFR) of the
magnet allows discrimination between muons and pions.

We reconstruct $\KS\to\pip\pim$
candidates from pairs of oppositely charged tracks. The two-track
combinations must form a vertex with $\pip\pim$ invariant mass within
\unit[$11.2$]{\mevcc} ($3.5\sigma$) of the etsablished \KS\ mass~\cite{pdg} and
reconstructed proper lifetime greater than five times its uncertainty.
We form $\piz\to\gamma\gamma$ candidates from pairs of photon
candidates in the EMC that are isolated from any charged tracks, carry
a minimum energy of \unit[50]{\mev}, fall within the mass window
\unit[$110<m_{\gamma\gamma}<160$]{\mevcc}, and produce the expected lateral
shower shapes.  Finally, we construct \Bztokspiz{} candidates by
combining \KS{} and \piz{} candidates in the event.  For each $B$
candidate two nearly independent kinematic variables are computed.
The first one is $\mb$, the invariant mass of the reconstructed $B$ meson, \Brec{}.
The second one is $\mmiss$, the invariant mass of the other $B$, \Btag{}, 
computed from the known beam energy, applying a mass constraint to \Brec{}.
For signal decays, the
two variables peak near the \Bz{} mass with a resolution of
\unit[$\sim5.5$]{\mevcc} (\unit{$\sim 31$}{\mevcc}) for $\mmiss$ ($\mb$). 
Both the $\mmiss$ and $\mb$ distributions exhibit a low-side tail from 
leakage of energy deposits out of the EMC.  We select candidates within the window
\unit[$5.11<\mmiss<5.31$]{\gevcc} and \unit[$5.1294<\mb<5.4294$]{\gevcc},
which includes the signal peak and a ``sideband'' region for
background characterization. For the~\unit[$0.8$]{\%} of events
with more than one candidate, we select the combination with the
smallest $\chi^2=\sum_{i=\piz,\KS} (m_i-m'_i)^2/\sigma^2_{m_i}$, where
$m_i$ ($m'_i$) is the measured (etablished) mass and $\sigma_{m_i}$ is
the estimated uncertainty on the measured mass of particle $i$.

The sample of \Bztokspiz{} candidates is dominated by random $\KS\piz$
combinations from $\epem\to\qqbar$ $(\q=\{u,d,s,c\})$ fragmentation.
Using large samples of simulated \BB{} events, we find that backgrounds
from other \B{} meson decays can be neglected. We exploit topological observables to
discriminate the jet-like $\epem\to\qqbar$ events from the more
uniformly distributed \BB{} events. 
We compute the value of $L_2/L_0$, where
$L_j\equiv\sum_i |{\bf p}^*_i| |\cos \theta^*_i|^j$.
Here, ${\bf p}^*_i$ is the momentum of particle $i$ in the $\Upsilon(4S)$ rest frame and
$\theta^*_i$ is the angle between ${\bf p}^*_i$ and the sphericity
axis~\cite{Bjorken:1969wi} of the \Bz{} candidate, and the sum does
not include the decay tree of the reconstructed $B$.
In order to reduce the number of background events, we require
$L_2/L_0<0.55$. We also use the distribution of this ratio to 
discriminate the signal from the residual background.
Using a full detector simulation, we estimate that
our selection retains $(34.3 \pm 1.3)\%$ of the signal
events. Here, the error includes statistical and systematic 
contributions. The systematic contribution is dominated by the
reconstruction of \KS\ and \piz{}.

For each \Bztokspiz{} candidate, we examine the remaining tracks and
neutral candidates in the event to determine if the \Btag{} meson
decayed as a \Bz{} or a \Bzb{} (flavor tag).  
We use a neural network (NN) to determine the flavor of the $\Btag$
meson from kinematic and particle identification
information~\cite{ref:sin2betaPRL02}. Each event is assigned to one of
seven mutually exclusive tagging categories, designed to combine flavor
tags with similar performance and vertex resolution.  We
parameterize the performance of this algorithm in a data sample
($B_{\rm flav}$) of fully reconstructed $\Bz\to D^{(*)-}
\pip/\rho^+/a_1^+$ decays. The average effective tagging efficiency
obtained from this sample is $Q = \sum_c \epsilon_S^c
(1-2w^c)^2=(30.4 \pm 0.3)\%$, where $\epsilon_S^c$ and $w^c$ are the
efficiencies and mistag probabilities, respectively, for events tagged
in category $c$. We take into account differences in tagging 
efficiency (for signal and background) and mistag (only for signal)
for $\Bz$ and $\Bzb$ events, in order to exclude
any source of fake CPV effects.
For the background, the fraction of events
($\epsilon_B^c$) and the asymmetry in the rate of $\Bz$ versus $\Bzb$
tags in each tagging category are extracted from the fit to the data.

Time-dependent
CPV asymmetries are determined by reconstructing the distribution of
the difference of the proper decay times, $\deltat\equiv
t_{\CP}-t_\text{tag}$, where the $t_{\CP}$ refers to the signal \Bz{}
and $t_\text{tag}$ to the \Btag{}. At the $\Upsilon(4S)$ resonance,
the $\deltat$ distribution follows
\begin{eqnarray}
  \label{eqn:td} 
  {\cal P}^{\Bz}_{\Bzb}(\deltat) \; = \; \frac{e^{-|\deltat|/\tau}}{4\tau} \times 
   \left[ \: 1 \; \pm \; 
    \left( \: S_f \sin{( \deltamd\deltat)} - C_f \cos{( \deltamd\deltat)} \: \right) \: \right] \; , 
\end{eqnarray}
where the upper (lower) sign corresponds to \Btag{} decaying as \Bz{}
(\Bzb), $\tau$ is the \Bz{} lifetime averaged over the two mass
eigenstates, \deltamd{} is the mixing frequency, $C_f$ is the
magnitude of direct CP violation in the decay to final state $f$, and
$S_f$ is the magnitude of CP violation in the interference between
mixing and decay. For the case of pure penguin dominance, we expect
$S_{\KS\piz}=\sin2\beta$, and $C_{\KS\piz}=0$.

We compute the proper time difference \deltat{} from the known boost
of the \epem{} system and the measured $\deltaz=\zrec-\ztag$, the
difference of the reconstructed decay vertex positions of the
\Bztokspiz{} and \Btag{} candidate along the boost direction ($z$).  A
description of the inclusive reconstruction of the \Btag{} vertex is
given in \cite{ref:Sin2betaPRD}.  For the \Bztokspiz{} decay, where no
charged particles are present at the decay vertex, we identify the
vertex of the fully reconstructed $B$ using
the single \KS{} trajectory from the $\pi^+ \pi^-$ momenta 
and the knowledge of the average
interaction point (IP), which is determined on a run-by-run basis from
the spatial distribution of vertices from two-track events.  We
compute \deltat{} and its uncertainty from a geometric fit to the
$\Upsilon(4S)\to\Bz\Bzb$ system that takes this IP constraint into
account. We further improve the sensitivity to \deltat{} by
constraining the sum of the two $B$ decay times
($t_{\CP}+t_\text{tag}$) to be equal to $2\:\tau$ with an
uncertainty $\sqrt{2}\; \tau_{\Bz}$, which effectively constrains the
two vertices to be near the \Y4S{} line of flight. We have verified in
a Monte Carlo simulation that this procedure provides an unbiased
estimate of \deltat{}.

The per-event estimate of the uncertainty on \deltat{} reflects the
strong dependence of the \deltat{} resolution on the $\KS$ flight
direction and on the number of SVT layers traversed by the $\KS$ decay
daughters. In about \unit[$60$]{\%} of the events, both pion tracks 
are reconstructed from at least 4 SVT hits, leading to sufficient resolution
for the time-dependent measurement. The average \deltat{} resolution
in these events is about \unit[$1.0$]{ps}. For events which fail this
criterion or for which \unit[$\sigma_{\deltat}>2.5$]{ps} or
\unit[$\deltat>20$]{ps}, the \deltat{} information is not used.
However, since \cf{} can also be extracted from flavor tagging
information alone, these events still contribute to the measurement of
\cf{} and the signal yield.

We obtain the probability density function (PDF) for the
time-dependence of signal decays from the
convolution of Eq.~\ref{eqn:td} with a resolution function ${\cal
  R}(\delta t \equiv \deltat -\deltat_{\rm true},\sigma_{\deltat})$,
where $\deltat_{\rm true}$ is the true value of $\deltat$.
The resolution function is parameterized as the sum of a `core' and a
`tail' Gaussian, each with a width and mean proportional to the
reconstructed $\sigma_{\deltat}$, and a third Gaussian centered at
zero with a fixed width of \unit[$8$]{ps}~\cite{ref:Sin2betaPRD}.  We
have verified in simulation that the parameters of ${\cal R}(\delta t,
\sigma_{\deltat})$ for \Bztokspiz\ decays are similar to those
obtained from the $B_{\rm flav}$ sample, even though the distributions
of $\sigma_{\deltat}$ differ considerably. Therefore, we extract these
parameters from a fit to the $B_{\rm flav}$ sample.  We find that the
\deltat{} distribution of background candidates is well described by a
$\delta$ function convolved with a resolution function with the same
functional form as used for signal events. The parameters of the
background function are determined together with the CPV parameters
and the signal yield.

We extract the CPV parameters from an extended unbinned maximum-likelihood (ML) fit
to kinematic, event shape, flavor tag, and time structure variables.
We construct the likelihood from the product of
one-dimensional PDFs, since all the linear correlations are negligible. 
The systematic from residual correlations is taken into account, as
explained below.

The PDFs for signal
events are parameterized from either a largest sample of fully-reconstructed
$B$ decays in data or from simulated events.  
For background PDFs, we select the functional form from data in the
sideband regions, included in the fitted sample, of the other
observables where backgrounds dominate

The likelihood function is defined as:
{\small\begin{eqnarray*}
  {\cal L}(\sf,\cf,N_S,N_B,f_S,f_B,\vec{\alpha}) =\frac{e^{-(N_S+N_B)}}{(N_S+N_B)\,!}  \times  
  \prod_{i \in \mathrm{w/\,} \deltat}
  \left[ N_S f_S \epsilon^{c}_S{\cal P}_S(\vec{x}_i,\vec{y}_i;\sf,\cf) + 
    N_B f_B \epsilon^{c}_B {\cal P}_B(\vec{x}_i,\vec{y}_i;\vec{\alpha}) \right] \times  \nonumber\\ 
  \prod_{i \in  \mathrm{w/o\,} \deltat}
  \left[ N_S (1-f_S) \epsilon^{c}_S {\cal P}'_S(\vec{x}_i;\cf) + 
    N_B (1-f_B) \epsilon_B^{c} {\cal P}'_B(\vec{x}_i;\vec{\alpha}) \right],
\end{eqnarray*}}
\noindent{}where $f_S$ is the fraction of events with 
$\deltat$ information ($\mathrm{w/\,} \deltat$) and
$f_B$ is the fraction of events without it ($\mathrm{w/o\,} \deltat$). 

The probabilities ${\cal P}_S$ and ${\cal P}_B$ are
products of PDFs for signal ($S$) and background ($B$) hypotheses
evaluated for the measurements
$\vec{x}_i=\{\mb,\mmiss,L_{2}/L_{0},\costhetacms,\text{tag},\text{tagging
  category}\}$ and $\vec{y}_i=\{\deltat,\sigma_{\deltat}\}$. 
In the formula, $\vec{\alpha}$ represents the set of parameters that define
the shape of the  PDFs.
Along with the CPV
asymmetries \sf\ and \cf, the fit extracts the yields $N_S$ and $N_B$,
the fraction of events with $\deltat$ information $f_S$ and $f_B$,
and the parameters $\vec{\alpha}$ which describe the background PDFs.

Fitting the data sample of $17058$ $\Bztokspiz$ candidates, we find
\mbox{$N_S=425 \pm 28$} signal decays with \mbox{$\skspiz =
  0.33 \pm 0.26 \pm 0.04$} and \mbox{$\ckspiz =
  0.20 \pm 0.16 \pm 0.03$}, where the uncertainties are
statistical and systematic respectively. Taking into account the
selection efficiency and the number of \BB{} pairs 
included in the fitted data sample, we also obtain
${\cal B}(K^0\pi^0)= (10.5 \pm 0.7 \pm 0.5)\times 10^{-6}$. 

\begin{figure}[!tbp]
\begin{center}
\includegraphics[width=0.9\linewidth]{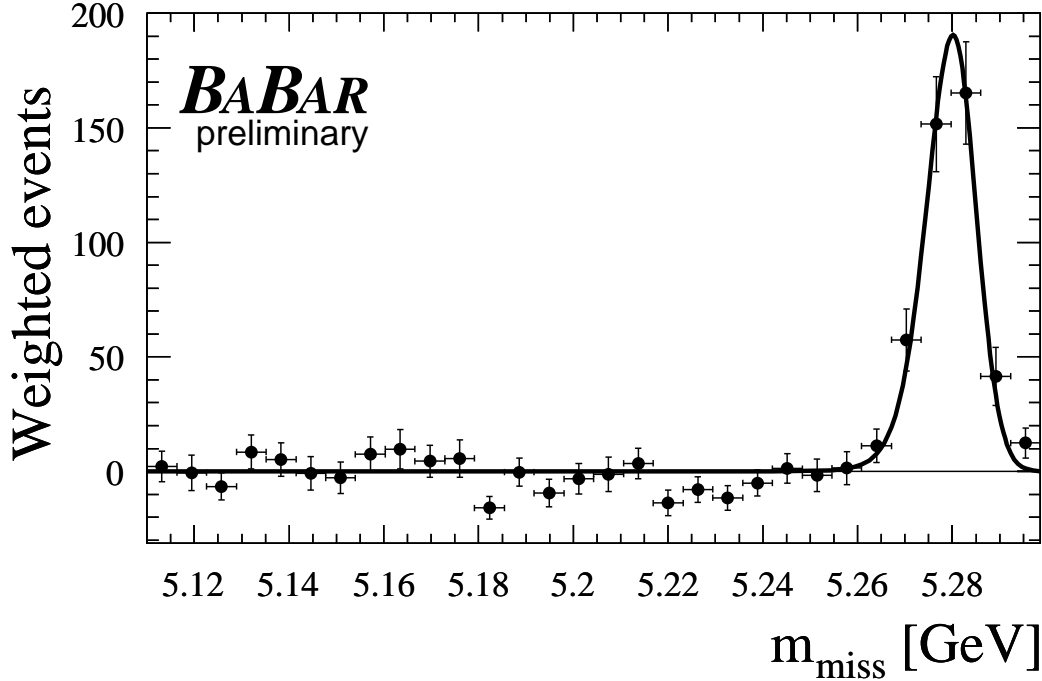}
\caption{$\mmiss$ distribution for signal events on data (dots), obtained using
the sPlot technique~\cite{splot} to subtract background events.
The solid curve represents the shape of signal PDF, as obtained from the fit.}
\label{fig:prplots}
\end{center}
\end{figure}

\begin{figure}[!tbp]
\begin{center}
\includegraphics[width=0.9\linewidth]{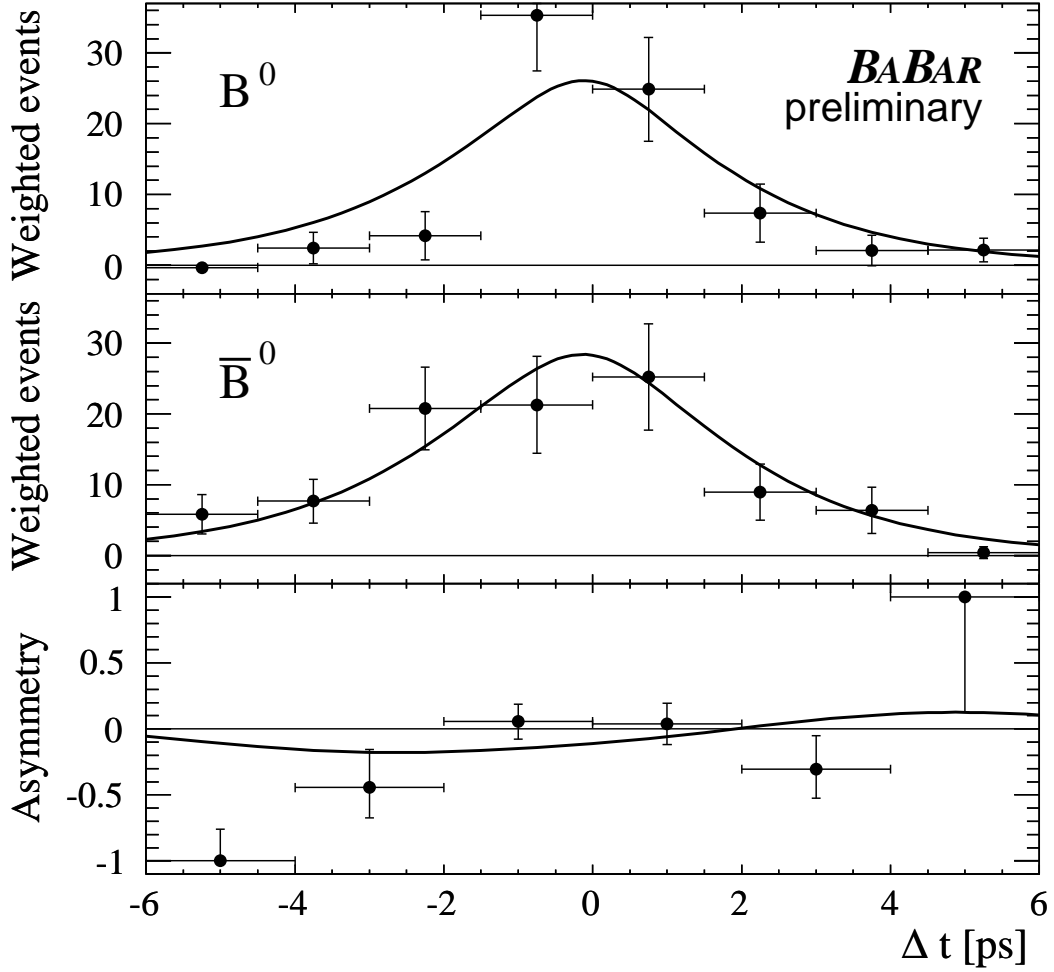}
\end{center}
\caption{Distributions of $\deltat$ for events weighted with the sPlot technique for
  $B_{\rm tag}$ tagged as (a) $\Bz$ or (b) $\Bzb$, and (c) the
  asymmetry ${\cal A}(\deltat)$.  The points are weighted data
  and the curves are the PDF projections. }
\label{fig:dtplot}
\end{figure}

Figure~\ref{fig:prplots} shows the $\mmiss$
distributions for signal events, where background is subtracted using
the sPlot technique~\cite{splot}.
Figure~\ref{fig:dtplot} shows distributions of $\deltat$ for $\Bz$-
and $\Bzb$-tagged events, and the asymmetry ${\cal
  A}_{\KS\piz}(\deltat) = \left[N_{\Bz} -
  N_{\Bzb}\right]/\left[N_{\Bz} + N_{\Bzb}\right]$ as a function of
$\deltat$, also obtained with the sPlot event weighting technique.
$N_\Bz$ ($N_{\Bzb}$) represents the number of events tagged as $\Bz$
($\Bzb$).

In order to investigate possible biases introduced in the CPV
measurements by the IP-constrained vertexing technique, we examine
\Bztojpsiks{} decays in data, where $\jpsi\to\mup\mun$ or $\jpsi\to
\epem$.  In these events we determine $\deltat$ in two ways: by fully
reconstructing the $\Bz$ decay vertex using the trajectories of
charged daughters of the $\jpsi$ and the $\KS$ mesons, or by
neglecting the $\jpsi$ contribution to the decay vertex and using the
IP constraint and the \KS{} trajectory only. This study shows that
within statistical uncertainties, the IP-constrained $\deltat$
measurement is unbiased with respect to the standard technique
and that the obtained values of $S_{\jpsi\KS}$ and $C_{\jpsi\KS}$ are
consistent. 

To compute the systematic error associated with the signal yield and CPV 
parameters, each of the input parameters to the likelihood fit is shifted
by $\pm 1\sigma$ from its nominal value and the fit is repeated.
Here, $\pm 1 \sigma$ is the associated error, as obtained from
the \Bflav{} sample (for $\Delta t$ and tagging) or from Monte Carlo.
This contribution to the systematic error takes into account the limited
statistics we used to parameterize the shape of the likelihood.
We obtain a systematic error of $0.72$ events on the yield, and
of $0.006$ ($0.010$) on \skspiz{} (\ckspiz{}).
As an additional systematic error associated with the shape of the
PDF, we also quote the largest deviation observed when the individual signal
PDFs are floated in the fit. This gives a systematic error of $11$ events on the yield, and
of $0.007$ ($0.021$) on \skspiz{} (\ckspiz{}).
The output values of the PDF parameters are also used to associate a systematic error to the 
selection cuts on the likelihood variables.
We evaluate the systematic error
coming from the neglected correlations among fit variables using a
set of toy Monte Carlo experiments, in which we embed signal events from
full detector simulations. We use the average shift in yield ($2.3$ events) and CPV
parameters ($0.003$ on \skspiz{} and $0.015$ on \ckspiz{}) as the associated uncertainty. 
We estimate the background from other $B$ decays to be negligible in
the nominal fit. We take into account a systematic error induced on signal yield and CPV
parameters by this neglected component, embedding $B$ background events
in the dataset and evaluating the  average shift in the fit result: 
$4.5$ events on the signal yield, 
$0.003$ on \skspiz{} and $0.002$ on \ckspiz{}.

For CPV parameters, we evaluate the additional systematic uncertainty related 
to the fit method using the largest 
difference between the fitted and generated values of \skspiz{} ($0.027$) and \ckspiz{} ($0.003$).
To quantify possible additional systematic effects, 
we examine large samples of simulated \Bztokspiz\ and \Bztojpsiks\ decays. 
We employ the
difference in resolution function parameters extracted from these
samples to evaluate uncertainties due to the use of the resolution
function ${\cal R}$ extracted from the $B_{\rm flav}$ sample. We
assign a systematic uncertainty of $0.01$ on \skspiz{} and $0.02$ on
\ckspiz{} due to the uncertainty in ${\cal R}$. 
We include a systematic uncertainty of $0.002$ on \skspiz{} and $0.001$ on \ckspiz{} 
to account for a possible misalignment of the SVT. We consider large variations of the IP
position and resolution, which produce a  systematic uncertainty of 
$0.004$ on \skspiz{} and $0.001$ on \ckspiz{}. 
Additional contributions come from the error on the known $B^0$ lifetime ($0.0022$ on both
\skspiz{} and \ckspiz{}), the value of $\Delta m_d$ ($0.0017$ on both
\skspiz{} and \ckspiz{}), and the effect of interference on the tag side ($0.0014$ on
\skspiz{} and $0.014$ on \ckspiz{}).

For the branching fraction, systematic errors come from the knowledge of
selection efficiency, $(34.3 \pm 1.3)\%$,
the counting of \BB{} pairs in the data 
sample, $(347.5 \pm 1.9)\times 10^{6}$ $\BB$ pairs, and the
branching fractions of the $B$ decay chain (${\cal B}(K^0_S \to \pi^+\pi^-)=0.6895 \pm 0.0014$
and ${\cal B}(\pi^0 \to \gamma \gamma) = 0.9880 \pm 0.0003$).~\cite{pdg}

In summary, we have performed a measurement of the time-dependent CPV
asymmetries of \Bztokspiz{} and the branching fraction of \Bztokzpiz.
We measured the parameters of CPV asymmetry $\ckspiz = 0.20 \pm 0.16 \pm 0.03$ and 
$\skspiz = 0.33 \pm 0.26 \pm 0.04$, and the branching fraction
${\cal B}(\Bztokzpiz)= (10.5 \pm 0.7 \pm 0.5)\times 10^{-6}$.
The first error is statistical and the second systematic.
All the results presented here are preliminary.

\par
We are grateful for the excellent luminosity and machine conditions
provided by our \pep2\ colleagues, 
and for the substantial dedicated effort from
the computing organizations that support \babar.
The collaborating institutions wish to thank 
SLAC for its support and kind hospitality. 
This work is supported by
DOE
and NSF (USA),
NSERC (Canada),
IHEP (China),
CEA and
CNRS-IN2P3
(France),
BMBF and DFG
(Germany),
INFN (Italy),
FOM (The Netherlands),
NFR (Norway),
MIST (Russia),
MEC (Spain), and
PPARC (United Kingdom). 
Individuals have received support from the
Marie Curie EIF (European Union) and
the A.~P.~Sloan Foundation.

\end{document}